\documentclass[fleqn,usenatbib,usedcolumn]{mnras}
\usepackage[british]{babel}             
\usepackage{newtxtext}                  
\usepackage[slantedGreek]{newtxmath}    
\usepackage[T1]{fontenc}                
\usepackage{graphicx}                   
\usepackage{rotating}
\usepackage{dcolumn}
\newcolumntype{d}{D{.}{.}{-1}}
\def\dhead#1{\multicolumn{1}{c}{#1}}
\def\twolines#1#2{$\kern-6pt\Big\{ {\textrm{#1\hfill}\atop\textrm{#2\hfill}}$}

\usepackage{multirow}
\usepackage{hyperref}
\graphicspath{{.}{./FIGS/}{..}}

\hypersetup{pdfauthor={I. H. Whittam, M. J. Jarvis, D. A. Green, I. Heywood and J. M. Riley},
               pdftitle={The prevalence of core emission in faint radio galaxies},
               pdfkeywords={galaxies: active, radio continuum: galaxies, catalogues, surveys},
               bookmarksnumbered=true}
\hypersetup{colorlinks=true,
            linkcolor=blue,
            citecolor=blue,
            filecolor=blue,
            urlcolor=blue}
\volume{{\rm in prep.}}
\title[The prevalence of core emission in faint radio galaxies]{The prevalence of core emission in faint radio galaxies in the SKA Simulated Skies}

\author[I.~H.~Whittam et al.]{I.~H.~Whittam$^{1}$\thanks{iwhittam@uwc.ac.za}, M.~J.~Jarvis$^{1,2}$, D.~A.~Green$^3$, I.~Heywood$^{4,5}$ and J.~M.~Riley$^3$\\
   $^{1}$Department of Physics and Astronomy, University of the Western Cape, Robert Sobukwe Road, Bellville 7535, South Africa\\
   $^{2}$Astrophysics, University of Oxford, Denys Wilkinson Building, Keble Road, Oxford, OX1 3RH, UK\\
   $^{3}$Astrophysics Group, Cavendish Laboratory, 19 J.~J.~Thomson Avenue, Cambridge CB3 0HE, UK\\
   $^{4}$CSIRO Astronomy and Space Science, P.O. Box 76, Epping, NSW 1710, Australia\\
   $^{5}$Department of Physics and Electronics, Rhodes University, PO Box 94, Grahamstown, 6140, South Africa\\}

\date{Accepted ---; received ---; in original form ---}

\pagerange{\pageref{firstpage}--\pageref{lastpage}}

\pubyear{2016}
\begin{document}


\label{firstpage}

\maketitle

\begin{abstract}
Empirical simulations based on extrapolations from well-established low-frequency ($< 5$~GHz) surveys fail to accurately model the faint, high frequency ($>10$~GHz) source population; they under-predict the number of observed sources by a factor of two below $S_{18~\rm GHz} = 10$~mJy and fail to reproduce the observed spectral index distribution. We suggest that this is because the faint radio galaxies are not modelled correctly in the simulations and show that by adding a flat-spectrum core component to the FRI sources in the SKA Simulated Skies, the observed 15-GHz source counts can be reproduced. We find that the observations are best matched by assuming that the fraction of the total 1.4-GHz flux density which originates from the core varies with 1.4-GHz luminosity; sources with 1.4-GHz luminosities $< 10^{25} \rm W \, Hz^{-1}$ require a core fraction $\sim 0.3$, while the more luminous sources require a much smaller core fraction of $5 \times 10^{-4}$. The low luminosity FRI sources with high core fractions which were not included in the original simulation may be equivalent to the compact `FR0' sources found in recent studies.
\end{abstract}


\section{Introduction}\label{section:intro}

Radio galaxies are some of the brightest objects in the radio sky and as such have been widely studied for several decades (e.g.\ \citealt{1974MNRAS.167P..31F,1980ARA&A..18..165M,1995PASP..107..803U}). Powerful radio galaxies mainly fall into two distinct morphological classes; the luminous edge-brightened Fanaroff and Riley type-II sources (FRII; \citealt{1974MNRAS.167P..31F}), and the more centrally-brightened less luminous FRI sources. The vast majority of the observed flux in these sources come from the large lobes and jets, which can extend for up to several Mpc (for an overview see \citealt{1991bja..book...52M}), whilst the emission from their cores typically only make up a few percent of the total emission. The core prominence differs between the two morphological classes, with FRI sources tending to have a higher fraction of their total flux density coming from the core \citep{1997MNRAS.284..541M,2014MNRAS.437.3405L}. Several studies suggest that there is a relationship between the core power and the total power of a source (e.g.\ \citealt{1990A&A...227..351D,1988A&A...199...73G}), with core fraction decreasing exponentially with increasing source luminosity. Physically, the emission from the core can be considered as a measure of the instantaneous jet power, while the extended emission traces the past jet activity. The contribution of the core to the total emission can therefore provide valuable information about the activity of the central engine and the duty cycle of a source.
In order to fully understand the role that cores play we need to study both individual sources and the statistical properties of larger samples across a range of redshifts and luminosities.

Recent studies of lower power ($L_{1.4~\rm GHz} < 10^{25} \rm W \, Hz^{-1}$) radio galaxies in the local universe ($z \sim 0.1$) have shown that the majority of these sources lack the powerful extended emission characteristic of the classical FRI and FRII sources, and have a higher proportion of their emission originating from their cores. For example, \citet{2014MNRAS.438..796S} found that more than two-thirds of the local radio galaxy population selected at 20~GHz were compact radio galaxies (with linear sizes less than $\sim 15$~kpc). \citet{2005MNRAS.362....9B} studied 2215 radio galaxies in both FIRST and SDSS with $L_{1.4~\rm GHz} \sim 10^{22} - 10^{26} \rm W\, Hz^{-1}$ and found that 80 per cent of the sample are unresolved in FIRST (therefore have angular sizes less than 5~arcsec, corresponding to $\sim 15$~kpc).  A pilot study of 12 of the \citeauthor{2005MNRAS.362....9B} sources using high-resolution VLA data at 1.4, 4.5 and 7.5~GHz by \citet{2015A&A...576A..38B} showed that 11 of the 12 sources studied show a compact morphology, with emission extending at most to $3$~kpc. These compact radio galaxies have been referred to as `FR0' sources. \citet{2016MNRAS.462.2122W} studied a sample of faint radio galaxies at higher redshifts ($z \sim 1$) selected from the Tenth Cambridge (10C; \citealt{2011MNRAS.415.2708D,2011MNRAS.415.2699F,2016MNRAS.457.1496W}) survey at 15.7 GHz. They used 610-MHz GMRT observations to investigate the morphology of the sample and found that the bulk of the sources are radio galaxies which lack extended emission on arcsec-scales, suggesting that this compact radio galaxy population may be significant at higher redshifts. There is therefore mounting evidence for a population of radio galaxies with a deficit of extended emission compared to classical FRI and FRII radio galaxies. Almost all of these compact radio galaxies have $L_{1.4~\rm GHz} < 10^{25} \rm W\, Hz^{-1}$ and they are most numerous at $L_{1.4~\rm GHz} < 10^{24} \rm W\, Hz^{-1}$, they therefore overlap classical FRI sources in luminosity.

The SKA Simulated Skies is a suite of simulations of the radio and sub-millimetre universe; one part is a semi-empirical simulation of the extra-galactic radio continuum sky developed by \citet{2008MNRAS.388.1335W,2010MNRAS.405..447W} which we will refer to as S$^3$. It covers a $20 \times 20$~deg$^2$ sky area out to a redshift of 20, containing 320 million radio components and extending down to a flux density limit of 10~nJy at 151, 610~MHz, 1.4, 4.86 and 18~GHz. It encapsulated our best understanding of the radio sky in 2010 and therefore serves two main purposes: it allows this understanding to be tested against newer observations, and it can be used to optimise the design of telescopes and radio surveys, such as the Square Kilometre Array. It is widely used by the community for both these purposes, see e.g.\ \citet{2015aska.confE..18J,2015aska.confE..83M,2015aska.confE..69S}.

Sources in S$^3$ are split into six distinct source types which are treated differently: radio-quiet AGN, radio-loud AGN (which are split into FRI and FRII sources and GHz-peak spectrum (GPS) sources), quiescent star-forming and star-bursting galaxies. The radio-loud AGN in the simulation are drawn from the 151 MHz luminosity function derived by \citet{2001MNRAS.322..536W}, which consists of low- and high-luminosity components whose luminosity function have different functional forms. These low- and high-luminosity components are crudely identified as FRI and FRII sources respectively. The morphologies and spectra of the two populations are then modelled differently and the morphological parameters are tuned to fit the observed source counts at $S > 1$~mJy over a range of frequencies from 151~MHz through to 18~GHz. A small number of GPS sources are included by introducing a spectral turnover in the relevant source spectra.
Throughout this paper, `FRI sources' refer to sources classified as FRI sources in the simulation (note that FR0 sources are not represented in this model).

At 18~GHz the majority of the simulated sources with $S_{18~\rm GHz} > 0.5$~mJy are radio-loud AGN, with FRI sources making up 71 per cent followed by FRII sources with 13 per cent, while the remaining source types only contribute a few per cent each. This is broadly consistent with recent work by \citet{2015MNRAS.453.4244W} who studied a sample of sources selected at 15.7~GHz from the 10C sample and found that $> 90$ out of a sample of 96 sources with $S_{15.7~\rm GHz} > 0.5$~mJy are radio galaxies, while star-forming galaxies make a negligible contribution to the population. (Although note that the majority of the radio galaxies in the \citealt{2015MNRAS.453.4244W} 10C sample seem to lack the extended emission typical of FRI/FRII sources, even in lower-frequency, 5~arcsec resolution imaging.)

One type of observation that can easily be compared to S$^3$ is a radio source count. The 1.4-GHz source counts derived from the simulation are consistent with the observed counts down to at least the $\muup$Jy-level, unsurprisingly as the model was originally set up to match the counts at this frequency. At higher frequencies (e.g.\ 18~GHz), however, S$^3$ fails to reproduce the observed counts. The 18-GHz counts derived from the simulation under-predict the counts from the Ninth Cambridge (9C; \citealt{2003MNRAS.342..915W,2010MNRAS.404.1005W}) and 10C surveys by a factor of two below 10~mJy. Other leading high-frequency models (e.g.\ \citealt{2005A&A...431..893D,2011A&A...533A..57T}) also significantly under-predict the number of sources at faint flux densities, indicating that the faint, high-frequency source population is not well characterised. As the cores of radio galaxies play a more significant role at higher frequencies due to their flat spectral shape, this suggests that the core emission, which is directly related to the instantaneous jet power, may be poorly modelled in the simulations of these populations.

The spectral index of a radio galaxy can provide information about the relative contributions of the core and the extended structure to the total emission; the extended lobes tend to have steep spectra ($\alpha > 0.5$) because the predominant emission mechanism is optically-thin synchrotron, while the core emission has a much flatter spectrum ($\alpha < 0.5$) due to the superposition of many synchrotron self-absorbed spectra from the base of the jet (e.g.\ \citealt{1979ApJ...232...34B}). Radio galaxies which are more core-dominated therefore tend to have flatter spectra than those dominated by extended emission.
\citet{2013MNRAS.429.2080W} compared the spectral index distributions of a sample of 10C sources with a comparable sample selected from the S$^3$ simulation and found that they were significantly different; there are essentially no sources in the simulated sample with $\alpha < 0.3$\footnote{The convention $S \propto \nu^{-\alpha}$ for flux density $S$, frequency $\nu$ and spectral index $\alpha$ is used throughout this work.} while 40 per cent of the 10C sample have $\alpha^{15.7~\rm GHz}_{0.61~\rm GHz} < 0.3$.  This is probably because the spectra of the FRI sources, which dominate the simulated sample and are assumed to have steep spectra, are not modelled correctly. In particular, their cores are known to be more dominant than is assumed in the simulation (e.g.\ \citealt{2014MNRAS.437.3405L}).  This is discussed in more detail in Section~\ref{section:source_mod}.

The failure to model the spectra of the FRI radio galaxies in S$^3$ implies that the structures of these sources are not well known. Spectral ageing, which occurs as the synchrotron electrons lose energy over time, causes the spectra of the extended lobes to steepen with time.

As the vast majority of the faint radio galaxies in the simulation are classified as FRI sources, in this work we probe how these sources are modelled in order to better understand this population. We focus on reproducing the statistical properties of the population at high frequencies ($\sim 15$ GHz) as this is where the simulation fails. In this work we introduce more significant cores into the FRI sources, as it is known that the core fraction ($S_{\rm core}/S_{\rm tot}$, where $S_{\rm core}$ is the flux density originating from the core and $S_{\rm tot}$ is the total flux density of the source) is higher than assumed in the simulation (e.g.\ \citealt{2015A&A...576A..38B}), to investigate the extent to which this improves the simulation by comparing with recent high-frequency observations. The radio data used to in this work are discussed in Section~\ref{section:radio_data}. The changes made to the FRI sources are described in Section~\ref{section:source_mod} and the results are discussed in Section~\ref{section:discussion}. The conclusions are then presented in Section~\ref{section:conclusions}.

\section{Radio data used}\label{section:radio_data}

The 10C survey is the deepest high-frequency ($\gtrsim 10$~GHz) radio survey to date, so is the most appropriate dataset to use to test any model or simulation of the source population making up the faint ($ \lesssim 50$~mJy), high-frequency radio sky. It was observed with the Arcminute Microkelvin Imager (AMI; Zwart et al., 2008) Large Array in Cambridge, UK, and consists of a total of 27~deg$^2$ complete to 1~mJy in 10 different fields and a further 12~deg$^2$ complete to 0.5~mJy contained within these fields. It has recently been extended down to 0.1~mJy in two fields \citep{2016MNRAS.457.1496W}. The properties of a complete sample of nearly 100 sources in one of the fields with extensive multi-wavelength data available have been studied in detail by \citet{2013MNRAS.429.2080W,2015MNRAS.453.4244W,2016MNRAS.462.2122W}, using lower-frequency radio data along with near- and far-infrared, optical and X-ray data. The key results of relevance to this work are that essentially all ($\geq 96$~per cent) of the sources are radio galaxies, and they have flatter spectra and are more numerous than the simulation predicts. 

The spectral indices of these sources were studied using 610~MHz GMRT data \citep{2008MNRAS.387.1037G,2010BASI...38..103G} and 1.4~GHz WSRT data \citep{2012rsri.confE..22G}, along with FIRST and NVSS (both at 1.4~GHz). The large resolution difference between the 10C survey (beam size 30~arcsec), and some of the other surveys means that care needs to be taken when calculating the spectral indices. To minimise the effects of these issues, contour maps of the sources in the higher-resolution surveys (e.g.\ the GMRT survey) are examined by eye and images are smoothed to a comparable resolution to the 10C survey. Full details of these methods used for each of the different surveys are given in Section 3.2 of \citet{2013MNRAS.429.2080W}.

The morphologies of the sources were investigated using 610~MHz GMRT data with a resolution of $6 \times 5$~arcsec \citep{2008MNRAS.387.1037G,2010BASI...38..103G}; the majority of the sources are compact of these scales, with 67 per cent classified as compact, 19 per cent classified as FRI or FRII sources and the remaining 14 per cent showing some signs of extended emission \citep{2016MNRAS.462.2122W}. 

The two main data products used in this paper to compare to the simulation are the 10C source counts, which are presented in \citet{2011MNRAS.415.2708D} and extended in \citet{2016MNRAS.457.1496W} and the spectral index distribution, which is derived and discussed in \citet{2013MNRAS.429.2080W} and extended in \citet{2017MNRAS.464.3357W}.

\section{Modifying the FRI sources}\label{section:source_mod}

In S$^3$, each FRI source is modelled as consisting of a compact core component and two extended lobes, based on observations of high-flux density FRIs. The core component is modelled as having some spectral curvature, meaning that while the core spectra are flat at low frequencies, between 1.4 and 18 GHz they are relatively steep, with a mean spectral index of 0.60; the extended emission is assumed to have a constant spectral index of 0.75. The contribution of the core to the total 1.4-GHz flux density is small, with a distribution which peaks at a core fraction, $x$, where $x = S_{\rm core}/S_{\rm tot}$ ($S_{\rm core}$ is the flux density originating from the core and $S_{\rm tot}$ is the total flux density of the source) of $2.5 \times 10^{-3}$. This means that the overall spectral shape of the FRI sources, which make up the majority of sources in the simulation with $S_{18~\rm GHz} = 1$~mJy, is steep with a mean spectral index of 0.73.

There is evidence that the core prominence of FRI sources is higher than assumed by the model; for example \citet{2015A&A...576A..38B} find a core fraction ranging from 0.05 to 0.9. There is also evidence that at higher frequencies ($\nu > 10$~GHz) the spectra of the lower power sources are flatter than assumed by the simulation.
\citet{2013MNRAS.429.2080W} studied the properties of a sample of 10C sources and showed that the spectra of sources selected at 15~GHz are much flatter than this, with a median spectral index between 610-MHz and 15.7-GHz of 0.36 for sources with $S_{15.7~\rm GHz} \sim 1$~mJy (see \citealt{2013MNRAS.429.2080W} for a full discussion of the spectral indices of the observed and simulated samples). We suspect that this discrepancy between S$^3$ and the observations arises because the cores of these lower power sources are in fact more dominant, and have flatter spectra, than the simulation assumes. 

\begin{figure}
\centerline{\includegraphics[width=\columnwidth]{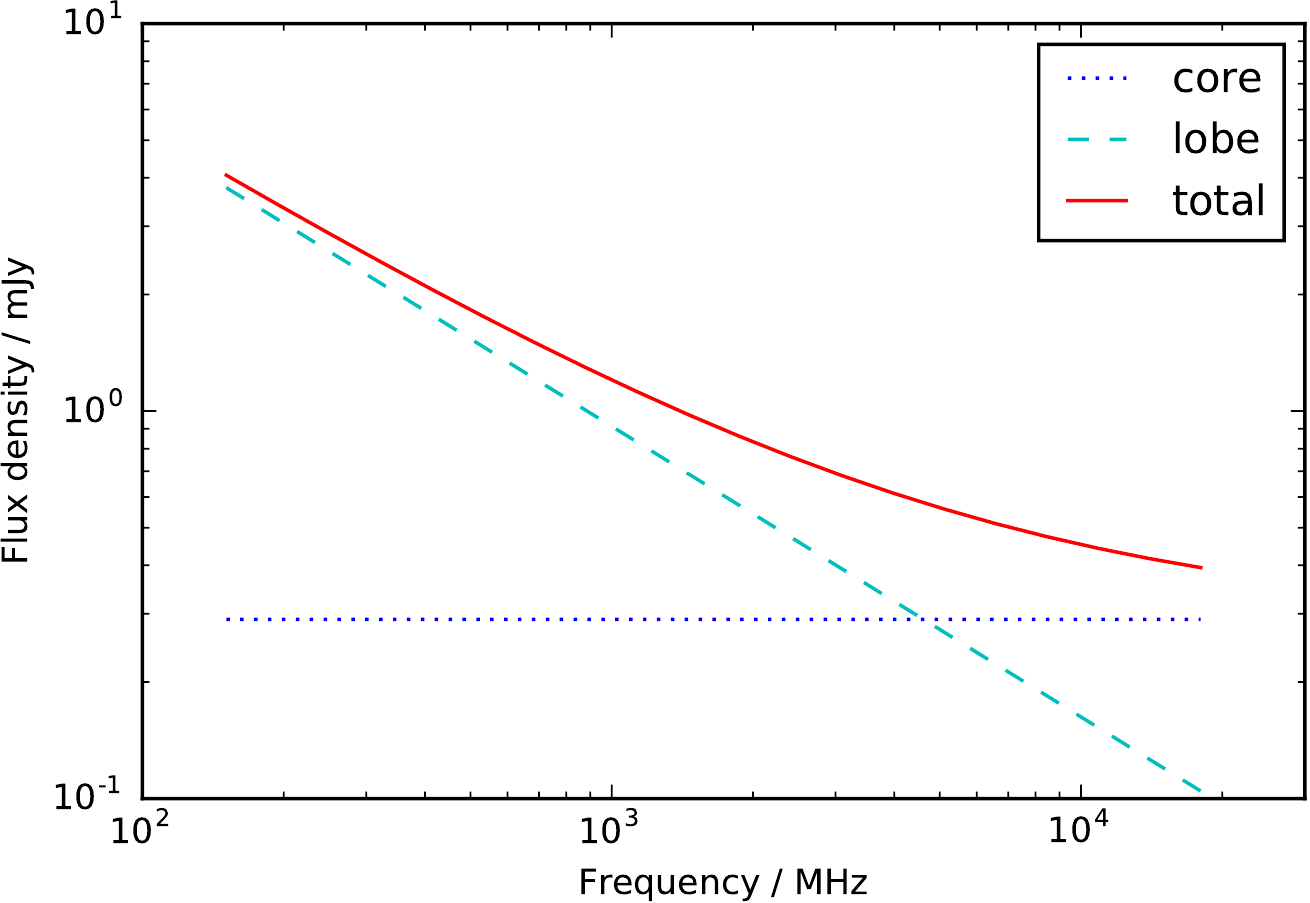}}
\caption{Example radio spectra of an FRI source in the modified catalogue. The source has $L_{1.4~\rm GHz} = 10^{24}~\rm W\, Hz^{-1}$ and a core fraction $x = 0.31$ at 1.4~GHz. The core and lobe components are shown separately, and have spectral indices of 0.0 and 0.75 respectively.}\label{fig:FRI_spectrum}
\end{figure}

\begin{figure}
\centerline{\includegraphics[width=\columnwidth]{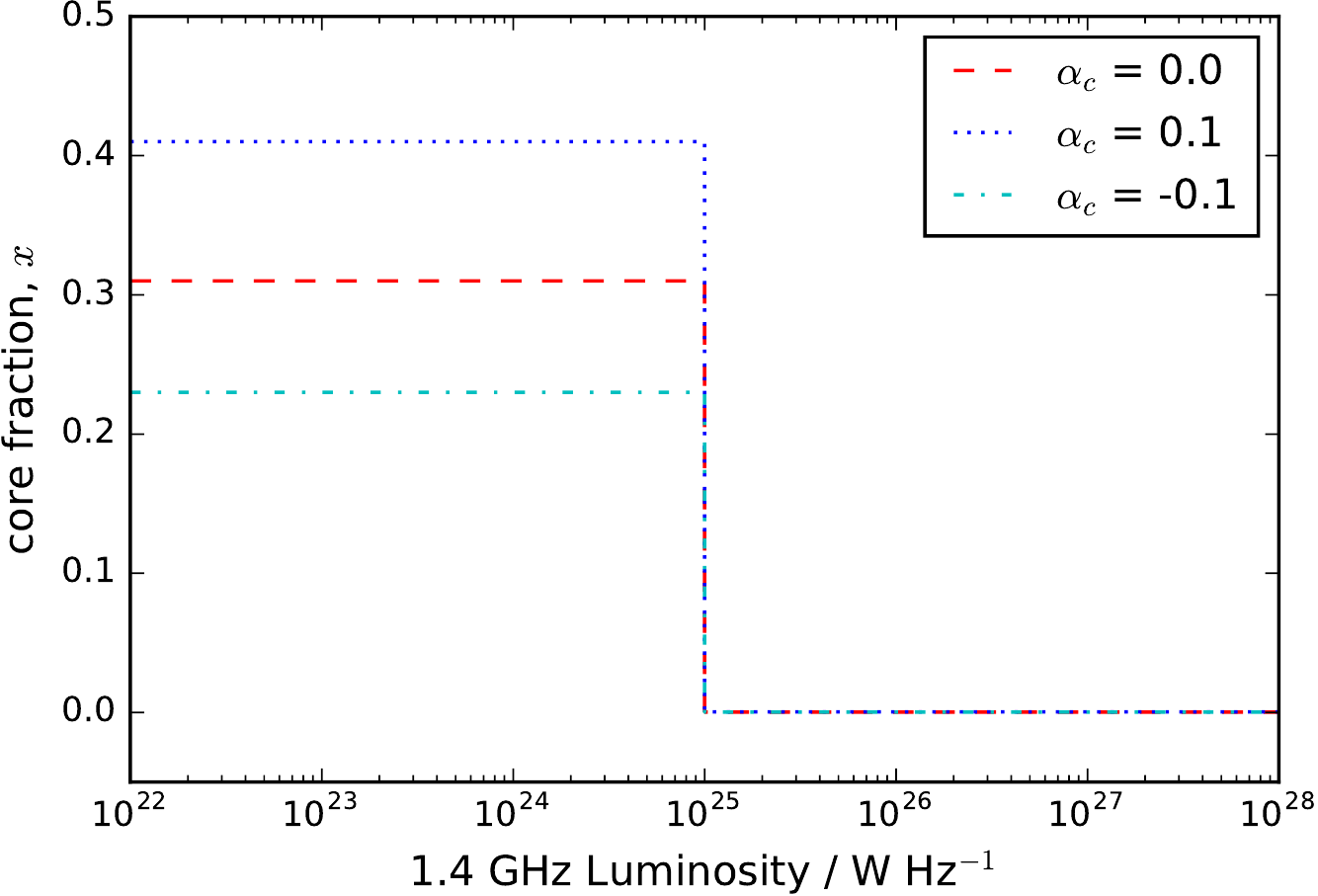}}
\caption{Fraction, $x$, of the total 1.4-GHz flux density attributed to the core as a function of 1.4-GHz luminosity for three different values of core spectral index.}\label{fig:core_frac}
\end{figure}

In order to test this theory, we have added a flat-spectrum core component to the FRI sources in S$^3$. We have assumed that a fraction $x$ of the total 1.4-GHz flux density is in the core, and assigned the core component a spectral index between 1.4 and 15.7~GHz of $\alpha_{\rm c}$. The extended emission, which accounts for the remainder of the flux density at 1.4 GHz, is given a spectral index of 0.75 (consistent with optically-thin synchrotron emission, in the same way as in the original simulation). (Note that, in reality, spectral ageing will cause the lobes of many sources to have steeper spectra at high frequencies. We make no attempts to include this effect here, leaving it for future work, but note that it would mean that our additional core components should be considered as lower limits as including this effect would cause a larger core component to be required. This is discussed in more detail in Section~\ref{section:discussion}.)
We then calculate a revised 15.7-GHz flux density for each FRI source, and use this to produce a revised 15.7-GHz source count. An example radio spectrum for a modified source is shown in Fig.~\ref{fig:FRI_spectrum}.

In order to fit the observed source counts across the full range of flux densities, a higher core fraction is required for the lower luminosity sources than for the higher luminosity sources. To achieve this we modelled the core fraction as a step function which depends on 1.4-GHz luminosity. We therefore had three parameters in our model: the core fraction required at low luminosities ($x_{\rm l}$), the core fraction required at high luminosities ($x_{\rm h}$), and the break in luminosity ($b$) where the change between the two core fractions occurs, as demonstrated in equation~\ref{eqn:core_frac}: 

\begin{equation}
x = \left\{
\begin{array}{llc}
x_{\rm l} & \textrm{for} & L_{1.4~\rm GHz} \leq b \\
x_{\rm h} & \textrm{for} & L_{1.4~\rm GHz} > b.\\
\end{array}\right.
\label{eqn:core_frac}
\end{equation}

\noindent We tried more complicated models where rather than a step function we had a gradually decreasing core component but found that the data did not warrant the inclusion of extra parameters. 

We allowed the three parameters in our core fraction model to vary in the following ranges: $1.0 \times 10^{-5} < x_{\rm h} < 0.1$, $0.1 < x_{\rm l} < 0.7$ and $1.0 \times 10^{22} < b < 1.0 \times 10^{28}$ and fix the core spectral index $\alpha_{\rm c}$. We then compared the simulated 15.7-GHz source count produced by each set of values to the observed source counts from the 9C, 10C and 10C ultra-deep surveys \citep{2010MNRAS.404.1005W,2011MNRAS.415.2708D,2016MNRAS.457.1496W} and used a simple chi-squared minimisation to select the best-fitting values. We repeated this process for different values of $\alpha_{\rm c}$ to produce a best-fit core fraction model for each value of core spectral index. The core spectral index and the core fraction are degenerate; the same result can be achieved by using a steeper core spectrum and a larger core fraction as by using a flatter spectrum and a smaller core fraction. To illustrate this, we found the best fit core fraction model using three different values of $\alpha_{\rm c}$; $-0.1$, 0.0 and 0.1. With more data it would be possible to investigate models in which the spectral index is related to the core fraction, but for the present study we just highlight this degeneracy. The core fraction found to best fit the observed counts in each case is tabulated in Table~\ref{tab:core_frac} and shown in Fig.~\ref{fig:core_frac}.

\begin{table}
\caption{Parameters of the core fraction function required to best fit the observed source counts assuming different core spectral indices ($\alpha_{\rm c}$). $x_{\rm l}$ and $x_{\rm h}$ are the core fractions at low and high luminosities respectively, and $b$ is the break in luminosities, as described by Equation~\ref{eqn:core_frac}.}\label{tab:core_frac}
\smallskip
\centering
\begin{tabular}{dccc}\hline
\dhead{$\alpha_{\rm c}$} & $x_{\rm l}$ & $x_{\rm h}$ & $b$ \\
 & & &(W / Hz) \\ \hline
-0.1 & 0.23 & $3.6 \times 10^{-4}$ & $1 \times 10^{25}$\\
0.0  & 0.31 & $4.8 \times 10^{-4}$ & $1 \times 10^{25}$\\
0.1  & 0.42 & $6.4 \times 10^{-4}$ & $1 \times 10^{25}$\\
\hline
\end{tabular}
\end{table}

\begin{figure}
\centerline{\includegraphics[width=\columnwidth]{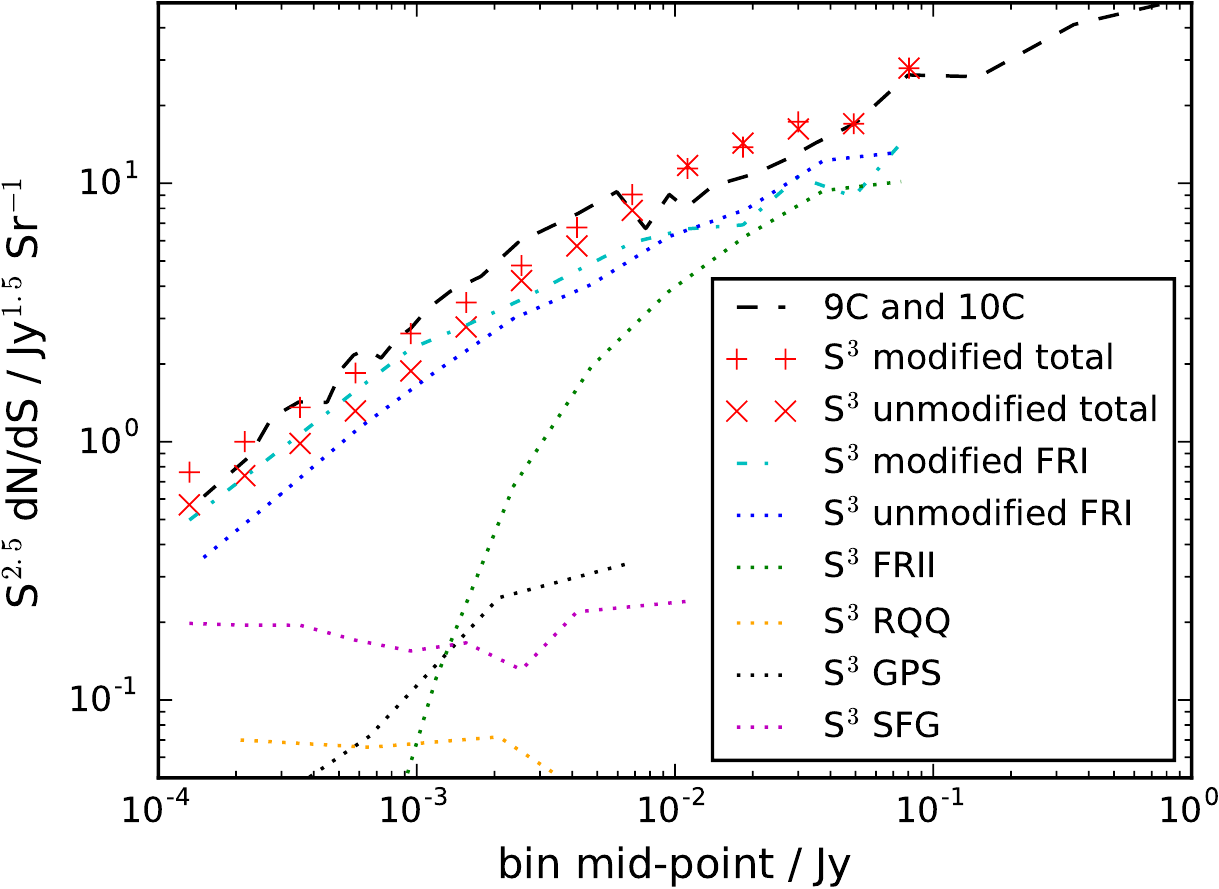}}
\caption{15.7~GHz Euclidean-normalised source counts. The red pluses ($+$) and crosses ($\times$) show the total source counts for the modified and original S$^3$ respectively. The dotted lines show the original S$^3$ counts split up into the different source populations used in the simulation, and the dot-dashed line shows the modified count for the FRI sources only (the only population modified here). The source counts from the 9C, 10C and 10C ultra-deep surveys are shown by the dashed line.}\label{fig:counts_18}
\end{figure}

\begin{figure}
\centerline{\includegraphics[width=\columnwidth]{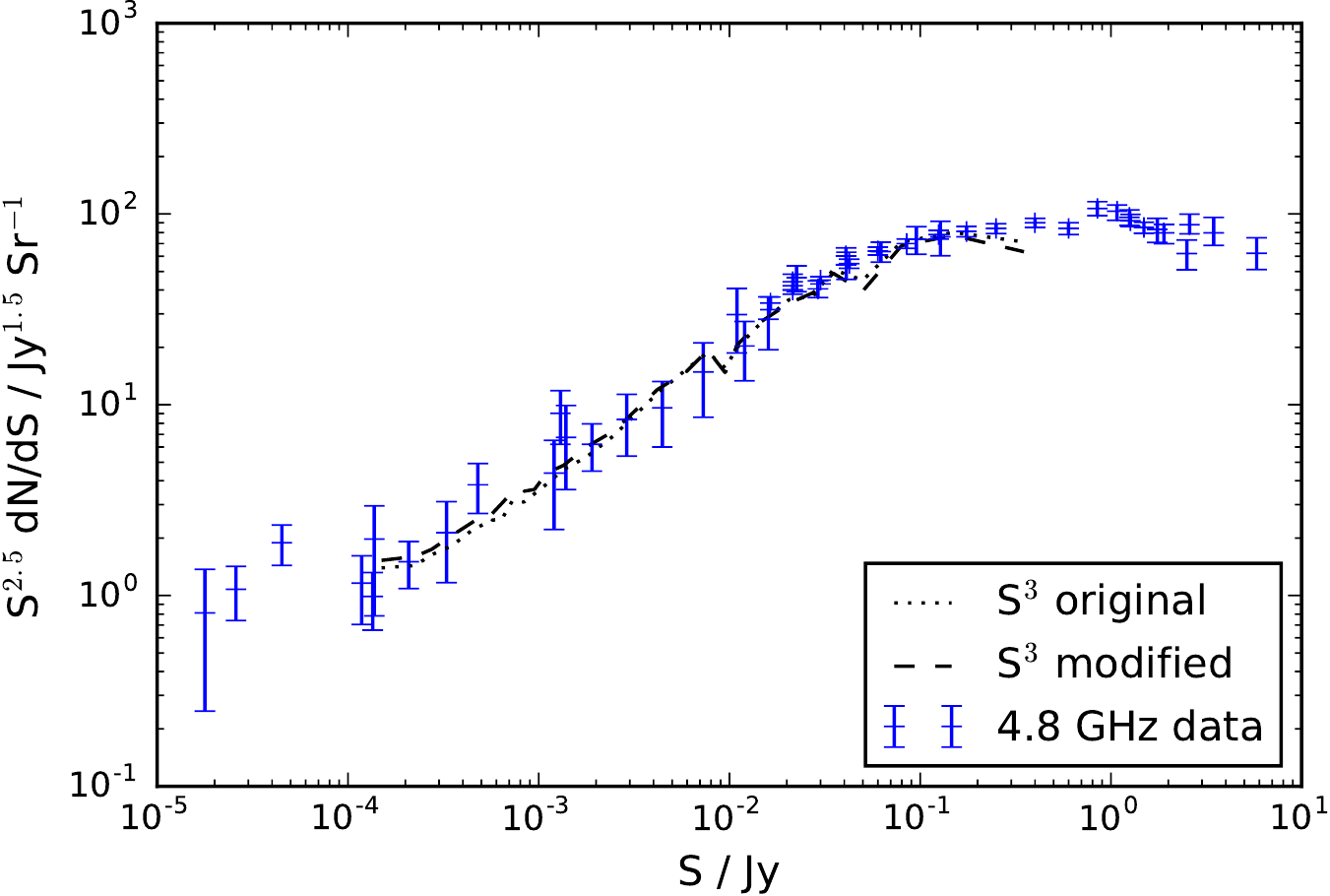}}
\caption{4.8-GHz source counts from the original and modified simulation. Observed counts, compiled from several different studies by \citet{2010A&ARv..18....1D} are also shown.}\label{fig:counts_4}
\end{figure}

The 15.7-GHz source counts resulting from the modified catalogue are shown in Fig.~\ref{fig:counts_18}, along with the observed counts from the 9C and 10C surveys and the original simulated counts. It is clear from this figure that the source counts from the modified simulation are a much better fit to the observed counts than those from the original catalogue.  The 4.8-GHz source counts, shown in Fig.~\ref{fig:counts_4}, are only slightly modified by the addition of the flat-spectrum cores, and are still consistent with the observations (compiled by \citealt{2010A&ARv..18....1D}).

\begin{figure}
\centerline{\includegraphics[width=\columnwidth]{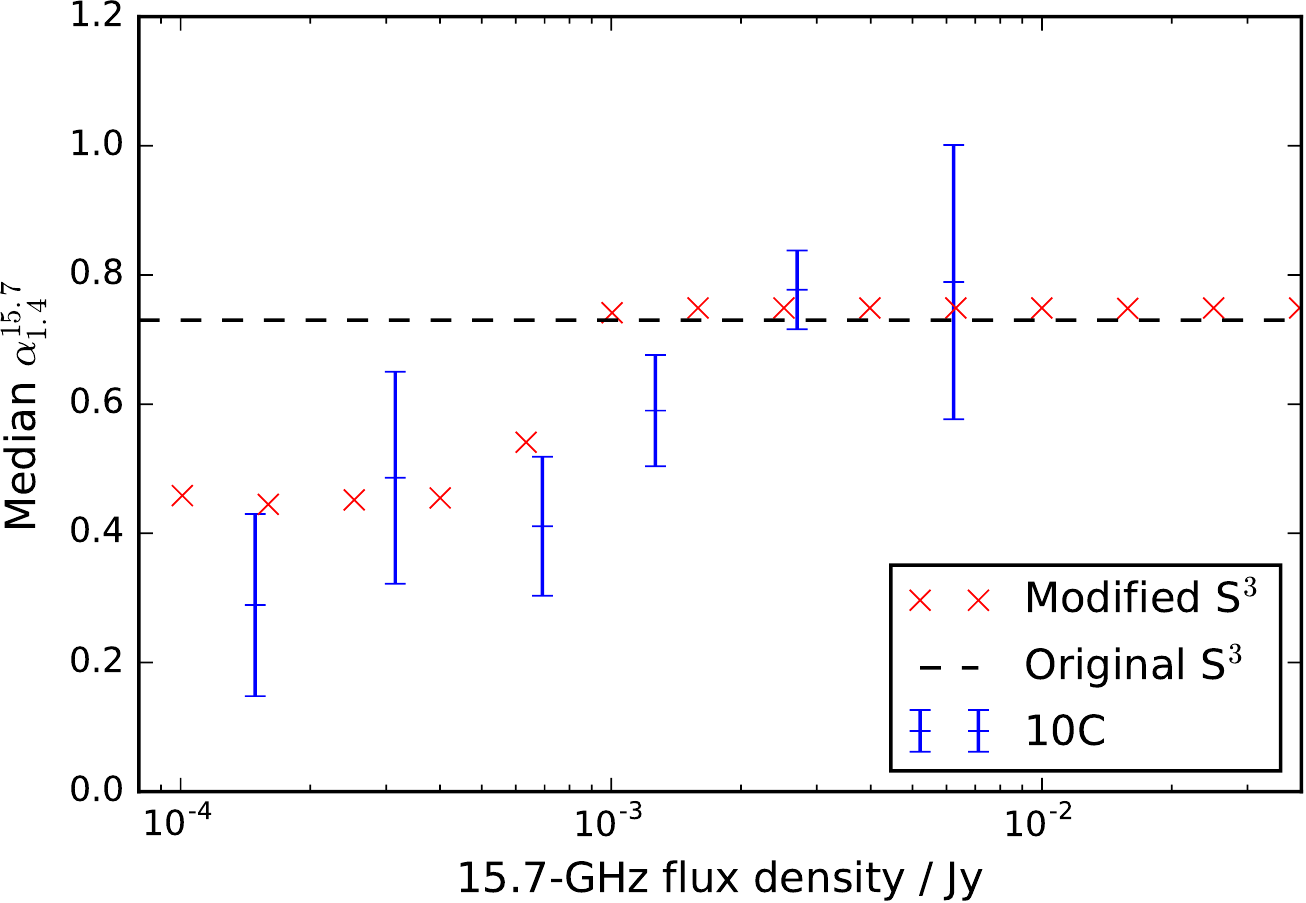}}
\caption{Median spectral index as a function of 15.7-GHz flux density. The median spectral index from the modified S$^3$ (assuming a core spectral index of 0.0) is shown as red `$\times$', while the observed median spectral index from the 10C and 10C ultra-deep surveys \citep{2013MNRAS.429.2080W,2017MNRAS.464.3357W} are shown in blue. }\label{fig:alpha}
\end{figure}

The median integrated spectral index as a function of flux density is shown in Fig.~\ref{fig:alpha}. In the original catalogue this was a particularly poor match to the observations; almost all sources in the simulation selected at 18~GHz had a spectral index of $\sim 0.75$, while the observed sample showed a much wider range of spectral indices. In particular, the observed sample contained a large proportion of flat-spectrum sources, which were totally missing from the simulation \citep{2013MNRAS.429.2080W,2017MNRAS.464.3357W}. The spectral index distribution from the modified simulated catalogue is clearly a much better match to the observations at 15~GHz.

\section{Discussion}\label{section:discussion}

The model presented in Section~\ref{section:source_mod} demonstrates one simple way in which S$^3$ can be altered to fit the observations; we have shown that by adding a flat-spectrum core component to the FRI sources with 1.4-GHz luminosities $ < 10^{25} \rm W \, Hz^{-1}$, the observed source counts and spectral index distributions can be reproduced.

As discussed in Section~\ref{section:intro}, there is growing evidence in the literature for a significant population of radio galaxies with $L_{1.4~\rm GHz} < 10^{25} \rm W \, Hz^{-1}$ which lack significant extended emission, both in the local universe \citep{2014MNRAS.438..796S} and at higher redshifts \citep{2016MNRAS.462.2122W}, particularly in samples selected at higher frequencies. This is consistent with our result that a significant core fraction is required in lower-luminosity sources to reproduce the observed high-frequency source count and spectral index distribution. The more core-dominated `FRI sources' in the modified simulation could well be equivalent to the compact (FR0) sources found in several recent surveys; the surface brightness of any extended emission could be too low to be detected in the observations, causing them to appear compact. \citet{2016MNRAS.tmp.1528S} modelled the emission from three VLBI-detected AGN and found that the predicted lobe surface brightness was too low to be detected in current surveys with the VLA and LOFAR, explaining why these sources appear compact. In fact, \citet{2015A&A...576A..38B} found that the only distinguishing factor between FR0 and FRI sources was the higher core dominance in FR0 sources; the two classes had very similar spectroscopic nuclear and host properties (e.g\ black hole mass, colour, emission line ratio).  

As shown in Fig.~\ref{fig:core_frac}, we have found that a significant core fraction is only required for sources with $L_{1.4~\rm GHz} < 10^{25} \rm W \, Hz^{-1}$; most of the FR0 sources in \citet{2014MNRAS.438..796S} and \citet{2015A&A...576A..38B} have luminosities below this value, so this is consistent with the scenario described above. Assuming a core spectral index of 0.0, we have found that a core fraction of 0.31 is required for the lower luminosity sources; the core dominance in the \citeauthor{2015A&A...576A..38B} FR0 sample ranges from 0.05 to 0.86, broadly consistent with the value found here, but the small sample size and large range makes it difficult to make a meaningful comparison. \citet{2014MNRAS.437.3405L} modelled the decelerating relativistic jets in ten low-luminosity ($L_{1.4~\rm GHz} \sim 10^{24} \rm W \, Hz^{-1}$) FRI sources selected from the 3CR and B2 samples. They found a range of core fractions, varying from 0.01 to 0.5. This indicates that a large spread of values are possible, depending on a variety of different factors, such as environment, jet speed and orientation angle with respect to the observer. 

For the powerful sources, with $L_{1.4~\rm GHz} > 10^{25} \rm W \, Hz^{-1}$, we have found that a core fraction of $5 \times 10^{-4}$ is required to match the observed source counts (again, assuming a core spectral index of 0.0). This is significantly smaller than has been found for bright sources in several observational studies. For example, \citet{2011MNRAS.417..184B} studied a sample of 31 FRI galaxies with $S_{5~\rm GHz} > 55$~mJy (and in many cases $S_{5~\rm GHz} > 1$~Jy) and found that the median ratio of the core emission to the total emission was $\sim 0.06$.  This could be because spectral ageing effects, which are not included in the original simulation or our modification, are significant for these brighter sources. Spectral ageing would cause the spectrum of the extended emission to steepen with increasing frequency, so including it would require a higher core fraction to be included to fit the observed source counts and spectral index distribution. For example, if spectral ageing were to increase the average spectral index between 1.4 and 15~GHz of the extended emission to 0.85 (rather than the 0.75 currently assumed for all extended emission), then a core fraction of 0.04 would be required.
Other studies, however, have found core fractions for brighter sources which are significantly lower. For example, \citet{1988A&A...199...73G} found a core fraction which decreased exponentially with increasing total radio power; this underestimates the core fraction that we have found to be required at luminosities less than $ 10^{25} \rm W \, Hz^{-1}$, predicting a core fraction of 0.02 at $L_{1.4~\rm GHz} = 10^{24} \rm W \, Hz^{-1}$, but is consistent with our results at higher luminosities, predicting a core fraction of 0.002 at $L_{1.4~\rm GHz} = 10^{26} \rm W \, Hz^{-1}$. \citet{1990A&A...227..351D} find a similar exponential relationship.
The abrupt change in core fraction used in this model is clearly not realistic, but we found that more complicated functions were not warranted but the data.

It is unclear what might cause some radio galaxies to have a paucity of extended emission. The core emission from a source is a measure of the instantaneous power produced by the jet, while the total emission traces the total jet power integrated over time. These lower luminosity radio galaxies with relatively high core fractions could therefore be young sources, or could be intermittent sources with a short duty cycle (e.g.\ \citealt{1994cers.conf...17R,2009ApJ...698..840C}). 
Alternatively, differences in the jet mechanism and/or environmental factors could be the cause.  \citet{2015A&A...576A..38B} suggest that FR0s have lower jet bulk speeds, $\Gamma$, than FRIs which cause their jets to be more subject to instabilities and entrainment, and therefore more likely to be prematurely disrupted. They speculate that a lower black hole spin may be the underlying cause of this lower jet speed, as a relationship between black hole spin and $\Gamma$ has been suggested by several authors \citep{2005ApJ...630L...5M,2012ApJ...759..114C}. \citet{2011MNRAS.417..184B}, however, find evidence that black hole spin is not as important as previously thought, and suggest that other factors, such as environmental density, the age of the radio sources and the black hole mass may play a more important role.

Indeed, there is evidence that the presence of a powerful core, indicating current AGN activity, is strongly dependent on the type of cluster in which a radio galaxy resides. \citet{2015MNRAS.453.1201H} studied a sample of 246 brightest cluster galaxies and found that a much higher proportion of `cool core' clusters hosted a radio galaxy with a powerful active core than `non-cool core' clusters; in their sample 60 per cent of `cool cores' compared to 12 per cent of `non-cool cores' have a distinguishable central engine, and 83 per cent of the `cool cores' have luminosities greater than $10^{23}~\rm W \, Hz^{-1}$ compared to just 5 per cent of non-cool cores. This suggests that the dominance of the core is strongly linked to the cluster state, so understanding the structures of faint radio galaxies is key to understanding how they relate to their environment.

Low-luminosity radio galaxies have not been studied in the way that the more luminous sources have, particularly at moderately high redshifts, and, as we have seen, models such as S$^3$ tend to assume they have less dominant cores than we have found necessary to reproduce the observed counts. 
The lower luminosity sources therefore have less dominant jets relative to their cores than previously estimated from observations at lower frequencies.   This has implications for the way in which radio sources are able to influence their host galaxy and surrounding environment -- understanding this is therefore key to understanding AGN feedback and the interplay between star-formation and AGN.
High resolution radio observations of a sample of high-frequency selected sub-mJy radio galaxies which allow their structure to be resolved are required to investigate this further.

\section{Conclusions}\label{section:conclusions}

We have shown that by adding a flat-spectrum core component to the FRI sources in S$^3$, the derived 18~GHz source counts, which were previously a poor fit to the observed counts below $S_{18~\rm GHz} \sim 10$~mJy, are consistent with the 9C and 10C source counts. The observed spectral index distribution, which was a particularly poor fit to the original simulation, is also reproduced. We find that the observed counts are well fitted by assuming that the sources have cores with a spectral index $\alpha^{15.7}_{1.4} = 0.0$ and fraction of the total 1.4-GHz flux density which increases with decreasing 1.4-GHz luminosity from $5 \times 10^{-4}$ for $L_{1.4~\rm GHz} > 10^{25} \rm W \, Hz^{-1}$ to 0.31 for $L_{1.4~\rm GHz} < 10^{25} \rm W \, Hz^{-1}$. 

These FRI sources with higher core fractions are therefore less dominated by extended emission, and may be equivalent to the compact FR0 sources found to dominate several recent surveys.

The physical interpretation of the paucity of extended emission in these radio galaxies is unclear; it could be due to a lower black hole spin, resulting in a lower jet speed, or differing jet lifetimes, or caused by environmental factors.

\section*{Acknowledgements}

We thank the referee for their comments on this manuscript. We thank Rob Fender for useful discussions. IHW and MJJ acknowledge the financial assistance of the South African SKA Project (SKA SA) towards this research (www.ska.ac.za). This research has made use of NASA's Astrophysics Data System. IHW thanks the South African Astronomical Observatory, where some of this work was carried out.

%
%

\setlength{\labelwidth}{0pt}

\bsp

\label{lastpage}

\begin{thebibliography}{}


 \bibitem[AMI Consortium: Davies et al.(2011)Davies et al.]{2011MNRAS.415.2708D} 
  AMI Consortium: Davies, et al., 2011, MNRAS, 415, 2708

 \bibitem[AMI Consortium: Franzen et al.(2011)Franzen et al.]{2011MNRAS.415.2699F} 
  AMI Consortium: Franzen, et al., 2011, MNRAS, 415, 2699 

 \bibitem[Baldi et al.(2015)Baldi, Capetti, \& Giovannini]{2015A&A...576A..38B} 
  Baldi R.~D., Capetti A., Giovannini G., 2015, A\&A, 576, A38 

 \bibitem[Best et al.(2005)Best et al.]{2005MNRAS.362....9B} 
  Best P.~N., Kauffmann G., Heckman T.~M., Ivezi{\'c} {\v Z}., 2005, MNRAS, 362, 9 

 \bibitem[Blandford \& K{\"o}nigl(1979)Blandford \& K{\"o}nigl]{1979ApJ...232...34B} 
  Blandford R.~D., K{\"o}nigl A., 1979, ApJ, 232, 34 

 \bibitem[Broderick \& Fender(2011)Broderick \& Fender]{2011MNRAS.417..184B} 
  Broderick J.~W., Fender R.~P., 2011, MNRAS, 417, 184 

 \bibitem[Chai et al.(2012)Chai, Cao, \& Gu]{2012ApJ...759..114C} 
  Chai B., Cao X., Gu M., 2012, ApJ, 759, 114 

 \bibitem[Czerny et al.(2009)Czerny et al.]{2009ApJ...698..840C} 
  Czerny B., Siemiginowska A., Janiuk A., Nikiel-Wroczy{\'n}ski B., Stawarz {\L}., 2009, ApJ, 698, 840 

 \bibitem[de Ruiter et al.(1990)de Ruiter et al.]{1990A&A...227..351D} 
  de Ruiter H.~R., Parma P., Fanti C., Fanti R., 1990, A\&A, 227, 351 

 \bibitem[de Zotti et al.(2005)de Zotti et al.]{2005A&A...431..893D} 
  de Zotti G., Ricci R., Mesa D., Silva L., Mazzotta P., Toffolatti L., Gonz{\'a}lez-Nuevo J., 2005, A\&A, 431, 893 

 \bibitem[de Zotti et al.(2010)de Zotti et al.]{2010A&ARv..18....1D} 
  de Zotti G., Massardi M., Negrello M., Wall J., 2010, A\&ARv, 18, 1 

 \bibitem[Fanaroff \& Riley(1974)Fanaroff \& Riley]{1974MNRAS.167P..31F} 
  Fanaroff B.~L., Riley J.~M., 1974, MNRAS, 167, 31P 

 \bibitem[Fender(2001)Fender]{2001MNRAS.322...31F} 
  Fender R.~P., 2001, MNRAS, 322, 31 

 \bibitem[Garn et al.(2008)Garn et al.]{2008MNRAS.387.1037G} 
  Garn T., Green D.~A., Riley J.~M., Alexander P., 2008, MNRAS, 387, 1037 

 \bibitem[Garn et al.(2010)Garn et al.]{2010BASI...38..103G} 
  Garn T.~S., Green D.~A., Riley J.~M., Alexander P., 2010, Bulletin of the Astronomical Society of India, 38, 103 

 \bibitem[Guglielmino et al.(2012)Guglielmino et al.]{2012rsri.confE..22G} 
 Guglielmino G., Prandoni I., Morganti R., Heald G., 2012, in `'Resolving The Sky -- Radio Interferometry: Past, Present and Future', available online at http://pos.sissa.it/cgi-bin/reader/conf.cgi?confid=163, id.22

 \bibitem[Giovannini et al.(1988)Giovannini et al.]{1988A&A...199...73G} 
  Giovannini G., Feretti L., Gregorini L., Parma P., 1988, A\&A, 199, 73 

 \bibitem[Heywood et al.(2016)Heywood et al.]{2016MNRAS.460.4433H} 
  Heywood I., et al., 2016, MNRAS, 460, 4433 

 \bibitem[Hogan et al.(2015)Hogan et al.]{2015MNRAS.453.1201H} 
  Hogan M.~T., et al., 2015, MNRAS, 453, 1201 


 \bibitem[Jarvis et al.(2015)Jarvis et al.]{2015aska.confE..18J} 
  Jarvis M., Bacon D., Blake C., Brown M., Lindsay S., Raccanelli A., Santos M., Schwarz D.~J., 2015, in Proceedings of Advancing Astrophysics with the Square Kilometre Array (AASKA14). 9 -13 June, 2014, Giardini Naxos, Italy. Online at http://pos.sissa.it/cgi-bin/reader/conf.cgi?confid=215, 18 

 \bibitem[Laing \& Bridle(2014)Laing \& Bridle]{2014MNRAS.437.3405L} 
  Laing R.~A., Bridle A.~H., 2014, MNRAS, 437, 3405 

 \bibitem[Lindsay et al.(2014)Lindsay et al.]{2014MNRAS.440.1527L} 
  Lindsay S.~N., et al., 2014, MNRAS, 440, 1527 

 \bibitem[Luchsinger et al.(2015)Luchsinger et al.]{2015AJ....150...87L} 
  Luchsinger K.~M., et al., 2015, AJ, 150, 87 

 \bibitem[McAlpine et al.(2015)McAlpine et al.]{2015aska.confE..83M} 
  McAlpine K., et al., 2015, in Proceedings of Advancing Astrophysics with the Square Kilometre Array (AASKA14). 9 -13 June, 2014, Giardini Naxos, Italy. Online at http://pos.sissa.it/cgi-bin/reader/conf.cgi?confid=215, 83 

 \bibitem[McKinney(2005)McKinney]{2005ApJ...630L...5M} 
  McKinney J.~C., 2005, ApJ, 630, L5 

 \bibitem[Miley(1980)Miley]{1980ARA&A..18..165M} 
  Miley G., 1980, ARA\&A, 18, 165 

 \bibitem[Morganti et al.(1997)Morganti et al.]{1997MNRAS.284..541M} 
  Morganti R., Oosterloo T.~A., Reynolds J.~E., Tadhunter C.~N., Migenes V., 1997, MNRAS, 284, 541 

 \bibitem[Muxlow \& Garrington(1991)Muxlow \& Garrington]{1991bja..book...52M} 
  Muxlow T.~W.~B., Garrington S.~T., 1991, Beams and Jets in Astrophysics. Edited by P.A. Hughes. Cambridge Astrophysics Series, No. 19. Cambridge, UK: Cambridge University Press, 1991., 52 

 \bibitem[O'Dea(1998)O'Dea]{1998PASP..110..493O} 
  O'Dea C.~P., 1998, PASP, 110, 493 

 \bibitem[Readhead et al.(1994)Readhead et al.]{1994cers.conf...17R} 
  Readhead A.~C.~S., Xu W., Pearson T.~J., Wilkinson P.~N., Polatidis A.~G., 1994, In Compact Extragalactic Radio Sources, eds. J. A. Zensus, \& K. I. Kellermann, 17 

 \bibitem[Sadler et al.(2014)Sadler et al.]{2014MNRAS.438..796S} 
  Sadler E.~M., Ekers R.~D., Mahony E.~K., Mauch T., Murphy T., 2014, MNRAS, 438, 796 

 \bibitem[Shabala et al.(2016)Shabala et al.]{2016MNRAS.tmp.1528S} 
  Shabala S.~S., Deller A., Kaviraj S., Middelberg E., Turner R.~J., Ting Y.~S., Allison J.~R., Davis T.~A., 2016, MNRAS, in press

 \bibitem[Simpson et al.(2012)Simpson et al.]{2012MNRAS.421.3060S} 
  Simpson C., et al., 2012, MNRAS, 421, 3060 

 \bibitem[Smolcic et al.(2015)Smolcic et al.]{2015aska.confE..69S} 
  Smolcic V., et al., 2015, in Proceedings of Advancing Astrophysics with the Square Kilometre Array (AASKA14). 9 -13 June, 2014, Giardini Naxos, Italy. Online at http://pos.sissa.it/cgi-bin/reader/conf.cgi?confid=215, 69 

 \bibitem[Snellen(2009)Snellen]{2009ASPC..402..221S} 
  Snellen I., 2009, in Approaching Micro-Arcsecond Resolution with VSOP-2: Astrophysics and Technologies ASP Conference Series, proceedings of the conference held 3-7 December, 2007, at ISAS/JAXA, Sagamihara, Kanagawa, Japan, eds. Yoshiaki Hagiwara, Ed Fomalont, Masato Tsuboi, and Yasuhiro Murata, 402, 221 

 \bibitem[Tucci et al.(2011)Tucci et al.]{2011A&A...533A..57T} 
  Tucci M., Toffolatti L., de Zotti G., Mart{\'{\i}}nez-Gonz{\'a}lez E., 2011, A\&A, 533, A57 

 \bibitem[Urry \& Padovani(1995)Urry \& Padovani]{1995PASP..107..803U} 
  Urry C.~M., Padovani P., 1995, PASP, 107, 803 

 \bibitem[Waldram et al.(2003)Waldram et al.]{2003MNRAS.342..915W} 
  Waldram E.~M., Pooley G.~G., Grainge K.~J.~B., Jones M.~E., Saunders R.~D.~E., Scott P.~F., Taylor A.~C., 2003, MNRAS, 342, 915 

 \bibitem[Waldram et al.(2010)Waldram et al.]{2010MNRAS.404.1005W} 
  Waldram E.~M., Pooley G.~G., Davies M.~L., Grainge K.~J.~B., Scott P.~F., 2010, MNRAS, 404, 1005 

  \bibitem[Whittam et al.(2013)Whittam et al.]{2013MNRAS.429.2080W} 
  Whittam I.~H., et al., 2013, MNRAS, 429, 2080 

 \bibitem[Whittam et al.(2015)Whittam et al.]{2015MNRAS.453.4244W} 
  Whittam I.~H., Riley J.~M., Green D.~A., Jarvis M.~J., Vaccari M., 2015, MNRAS, 453, 4244 

 \bibitem[Whittam et al.(2016a)Whittam et al.]{2016MNRAS.462.2122W} 
  Whittam I.~H., Riley J.~M., Green D.~A., Jarvis M.~J., 2016a, MNRAS, 462, 2122 

 \bibitem[Whittam et al.(2016b)Whittam et al.]{2016MNRAS.457.1496W} 
  Whittam I.~H., Riley J.~M., Green D.~A., Davies M.~L., Franzen T.~M.~O., Rumsey C., Schammel M.~P., Waldram E.~M., 2016b, MNRAS, 
457, 1496 

 \bibitem[Whittam et al.(2017)Whittam et al.]{2017MNRAS.464.3357W} 
  Whittam I.~H., Green D.~A., Jarvis M.~J., Riley J.~M., 2017, MNRAS, 464, 3357   

 \bibitem[Williams et al.(2016)Williams et al.]{2016MNRAS.460.2385W} 
  Williams W.~L., et al., 2016, MNRAS, 460, 2385 

 \bibitem[Willott et al.(2001)Willott et al.]{2001MNRAS.322..536W} 
  Willott C.~J., Rawlings S., Blundell K.~M., Lacy M., Eales S.~A., 2001, MNRAS, 322, 536 

 \bibitem[Wilman et al.(2008)Wilman et al.]{2008MNRAS.388.1335W} 
  Wilman R.~J., et al., 2008, MNRAS, 388, 1335 

 \bibitem[Wilman et al.(2010)Wilman et al.]{2010MNRAS.405..447W} 
  Wilman R.~J., Jarvis M.~J., Mauch T., Rawlings S., Hickey S., 2010, MNRAS, 405, 447 





\end{thebibliography}
\end{document}